\documentclass[conference]{IEEEtran}

\IEEEoverridecommandlockouts

\usepackage[utf8]{inputenc}

\usepackage[english]{babel}
\usepackage[left=3cm,right=2cm,top=2cm,bottom=1.9cm]{geometry} 
\usepackage[T1]{fontenc}
\usepackage{fancyhdr}
\usepackage[Bjornstrup]{fncychap}
\ChNumVar{\raggedright\fontsize{76}{80}\usefont{OT1}{pzc}{m}{n}\selectfont}
\ChTitleVar{\raggedleft\Large\sffamily\bfseries}

\usepackage{amsmath}
\usepackage{bm}
\usepackage{graphicx}
\usepackage{subcaption}

\usepackage{pdfpages}
\usepackage{comment}
\usepackage{multirow}
\usepackage{tabularx}
\usepackage{longtable}
\usepackage{fancybox}
\usepackage{lmodern}
\usepackage{amsmath}
\usepackage{mathtools}
\usepackage{breqn}
\usepackage[font=small,labelfont=bf]{caption}
\usepackage{enumitem}
\usepackage{textcomp}
\usepackage{color}
\usepackage[dvipsnames]{xcolor}
\usepackage[hidelinks]{hyperref}
\usepackage[nameinlink,capitalise]{cleveref}
\usepackage{multicol}
\usepackage{array}
\usepackage{float}
\usepackage{multirow}
\usepackage{gensymb}
\usepackage{url}
\usepackage[colorinlistoftodos,prependcaption]{todonotes}
\usepackage{siunitx,physics}
\usepackage{listings}
\usepackage{times}
\usepackage{verbatim}
\usepackage{amssymb}
\usepackage{mathtools}
\usepackage{siunitx}

\newcolumntype{P}[1]{>{\centering\arraybackslash}p{#1}}

\addto\captionsenglish{}

\usepackage{subfiles}
\usepackage{titlesec}
\usepackage{etoolbox}

\makeatletter
\patchcmd{\DOTI}{\vskip 30\p@}{\vskip 25\p@}{}{}
\patchcmd{\DOTIS}{\vskip 40\p@}{\vskip 5\p@}{}{}
\makeatother

\usepackage{cite}
\usepackage{color}
\usepackage{enumerate} 
\usepackage{amsmath,amssymb,amsfonts}
\usepackage{algorithmic}
\usepackage{graphicx}
\usepackage{multicol, blindtext}

\usepackage{textcomp}

\def\BibTeX{{\rm B\kern-.05em{\sc i\kern-.025em b}\kern-.08em
    T\kern-.1667em\lower.7ex\hbox{E}\kern-.125emX}}
\begin{document}

\title{Performance assessment of Synchronous Condensers vs Voltage Source Converters providing grid-forming functions \\
\thanks{
}
}

\author{
\IEEEauthorblockN{Dorsan Lepour \\ Mario Paolone}
\IEEEauthorblockA{\textit{École polytechnique fédérale de Lausanne} \\
Lausanne, Switzerland \\
dorsan.lepour@epfl.ch}
\and
\IEEEauthorblockN{Guillaume Denis \\ Carmen Cardozo \\ Thibault Prevost \\ Emeline Guiu}
\IEEEauthorblockA{\textit{Réseau de transport d'électricité} \\
Paris, France \\
guillaume.denis@rte-france.com}
}

\IEEEoverridecommandlockouts
\IEEEpubid{\makebox[\columnwidth]{978-1-6654-3597-0/21/\$31.00~\copyright2021 IEEE \hfill} \hspace{\columnsep}\makebox[\columnwidth]{ }}

\maketitle

\thispagestyle{plain}
\pagestyle{plain}

\vspace{-3em}
\begin{abstract}

Having sufficient grid-forming sources is one of the necessary conditions to guarantee the stability in a power system hosting a very large share of inverter-based generation. The grid-forming function has been historically fulfilled by synchronous machines. However, with the appropriate control, it can also be provided by voltage source converters (VSC). This work presents a comparison between two technologies with grid-forming capability: the VSC with a grid-forming control coupled with an adequate energy storage system, and the synchronous condensers (SC). Both devices are compared regarding their inertial response, as well as their contribution to the system strength and short-circuit current for an equivalent capacity expressed in terms of apparent power and inertial reserve. 
Their behaviour following grid disturbances is assessed through time-domain simulations based on detailed electromagnetic transient (EMT) models. The results show that both devices achieve similar performance in the time-scale of seconds. For shorter time-windows, however, they present a different behavior: the SC ensures a better stiffness in the first tens of ms following the disturbance, while the VSC offers a faster resynchronization.

\end{abstract}


\section{Introduction}
\label{introduction}
Power electronic converters used to interface renewable energy sources (RES) are intrinsically different from conventional synchronous generators (SG). Moreover, these converters have been historically controlled as current sources, such that the tracking of the power reference relies on the measurement of the grid voltage phasor inferred from its waveform using a fast phase locked loop in charge of the synchronization with the main grid. However, this approach has led to stability issues, especially in weak networks~\cite{Zhou2014PLL}. Power generation based on this type of converters is then referred to as grid-following or grid-feeding \cite{PSCC2020Palonoeetal}.

However, a power system cannot operate solely with grid-following converters, as the voltage would no longer be regulated by any source. In practice, stability limits have been reached at different RESs penetration levels, mainly in electrical islands such as South Australia~\cite{southaustraliaSC} and Texas~\cite{panhandle}, where the replacement of SG by RESs has led to a reduction of the system inertia and its strength. The consequences include the violation of security criteria and the curtailment of RES. As a short-term countermeasure, transmission systems operators (TSO) have been relying on synchronous condensers (SC) since they represent a mature technology, well known by operators and commercially available~\cite{Marken2011SCGE}. 

As known, SC are based on synchronous machines and, therefore, offer at least the same levers of conventional generators to ensure short-term stability (mechanical inertia, voltage regulation, over excitation, overload capability,...). They were often sized based on reactive power requirements for voltage regulation. With the emerging scarcities of inertia and system strength, traditionally quantified by the short-circuit power, other parameters become relevant in the technical specification of potential solution as function of the local system needs. As a consequence, manufacturers have boosted the SC contribution to these system properties 
by adding a flywheel or through a tailor made electromagnetic design.

It is now acknowledged that voltage source converters (VSC), with a grid-forming control and associated to a suitable energy storage system, can provide at least inertial response and contribute to system strength. Moreover, pilot projects based on energy storage systems (ESS)~\cite{ABBWIW2019,OSMOSED32} and a wind farm~\cite{SiemensWIW2019} have proved the technical feasibility and economical viability of this solution. Nonetheless, some barriers to the deployment of grid-forming VSC remain. On the one hand, grid-forming converters are not currently compliant with Connection Network Codes (CNC) and, in general, producers lack incentive to include this capability into their assets. This absence of demand, together with missing standards, have discouraged manufacturers from making this technology commercially available fast enough. These issues are being addressed at the regulatory level~\cite{NGGFdraft2020}.

On the other hand, the complexity of the power system dynamic assessment makes it challenging to anticipate the overall system needs and estimate the underlying economical opportunity. Finally, the potential over cost related to design changes necessary to provide grid-forming capability, can be sensitive to the detailed specification of the function, as this might impose some overcurrent capability or energy storage constraints. 

The MIGRATE EU project has made significant contributions to this topic by providing a definition of the grid-forming function that includes both technologies~\cite{MigrateD36}, SC and VSC, and performing first studies on the deployment levels required to ensure power system stability~\cite{MigrateD34}. Based on these results, a macro economic study on the French future network (Scenario Watt 2035\cite{RTEBilanPrev2017}) concluded that the installation of grid-forming VSC or SC to ensure stability would lead to comparable long term costs~\cite{PrevostEPE2020}. 
However, those solutions might not provide exactly the same services. Depending on the system needs, the optimal deployment of grid-forming sources to ensure stability may be a mix of both technologies. 

This work discusses the technical equivalence between the grid supporting services provided by a grid-forming VSC or a SC, for the same sizing in both inertial reserve, defined here through the inertia constant ($H$ in MW$\cdot$s/MVA) and apparent power (in MVA). 
For this purpose, we adopt the MIGRATE project definition of the grid-forming capability, which implies a voltage source behaviour, as well as the compatibility and synchronization with other grid-forming sources, including SG and any device connected to the grid, such as grid-following converters and loads. 
The focus of this study is on comparing the SC and the grid-forming VSC dynamic responses during the first seconds after an event.

In this work, we differentiate the capability of providing voltage stiffness in normal operation from the fault current injection as proposed in
~\cite{MigrateD23}. Therefore, the comparison between the grid-forming VSC and the SC is based on: (i) the inertial response following a load variation, (ii) the system strength/impedance after a grid voltage variation, and (iii) the short-circuit current supplied during a three-phase fault, although many others ancillary services have been defined~\cite{Irena2019AS}. To carry out these analyzes, we perform time-domain simulations using detailed EMT models of each device. The modeling details are provided in section~\ref{sec:model}. In section~\ref{sec:simu} we compare the SC and VSC dynamic responses when inserted in a transmission grid and facing local or remote disturbances. Finally, section~\ref{sec:conclusion} discusses the intrinsic capabilities of each technology in terms of grid support.

\section{Modeling and test description}
\label{sec:model}

In this work we consider three-phase EMT \textit{SimPowerSystems} models of:
\begin{itemize}
    \item A synchronous machine and its automatic voltage control (AVR).
    \item An average grid-forming VSC model\footnote{As the main interest is on system-level transients.}.
\end{itemize}

In both cases, only the circuit losses (resistive components) are considered, while mechanical (for SC) and commutation (for VSC) losses are neglected as they have no impact on the phenomena of interest. The parameters defining the equivalence for the comparison between the VSC and the SC are discussed in section~\ref{sec:eqsize}. In addition, both devices are connected to the same network through an identical step-up transformer. The grid is represented by a large equivalent machine instead of an infinite bus to take into account the interaction of the system frequency dynamics with the grid-forming device (see section~\ref{sec:grid}).

\subsection{Synchronous condenser model}

The SC model is built from a generic synchronous machine. It therefore provides an inherent inertial and damping response that depend on the inertia constant $H_{SC}$ and the damping factor $K_d$ as in~\eqref{eq:Hsc}: 
\begin{equation}
    \frac{d\omega}{dt} = \frac{1}{2H_{SC}}  \left( - \Delta P_e - K_d \left( \omega - \omega_g \right)  \right)
\label{eq:Hsc}
\end{equation}
where $\omega_g$ is the grid angular frequency, $\omega$ is the SC electrical angular speed, and $P_e$ is the electrical power provided to the system. The underlying hypothesis is that the rotor speed remains around its nominal value. A salient-pole rotor type unit with two pairs of poles is chosen as they are commonly used for this kind of application. The electromagnetic model (dq-axis transient, sub-transient reactances and time constants) is obtained from per unit quantities taken from a SC model EMT implementation realized by Hydro-Quebec \cite{reactances}. We consider an adaptation of the AC1A excitation system (IEEE 421 standard \cite{excitation}), consisting of an AC alternator driving a diode rectifier to produce the field voltage. It is controlled by a PI voltage regulator which takes as input a measurement of the stator voltage $v_s$ and outputs the field voltage $v_{fd}$ so as to meet the stator voltage reference $v_s^{*}$. The regulator gain and time constant were tuned manually in order to have a voltage regulation settling time of about 500 ms, which is a lower bound for conventional units with classical excitation system. 

\subsection{Voltage source converter model}

A VSC creates a sinusoidal voltage waveform autonomously, provided that it is supplied by a DC voltage source. It has then the capability to impose the amplitude and phase of the instantaneous AC voltage at its output, generating a stiff voltage waveform (i.e., not influenced by loading conditions). Here, we use the averaged two-level VSC model with LCL filter proposed in~\cite{GFMmodelurl}. The DC voltage source has been replaced by a model of a battery energy storage system (BESS), based on a capacitor $C_{dc}$ fed by a voltage controlled current source. The current $i_{dc,source}$ is controlled by a PI regulator taking as input the difference between the DC voltage $V_{dc}$ and its reference value $V_{dc}^{*}$. A symmetrical optimum method is used for the tuning of the controller gains, for a capacitor filtering time $t_{dc}$ chosen equal to 6 ms. However, the results remain valid for any other source capable of keeping a relatively stiff DC voltage~\cite{Cardozo2020SIW}. 

\begin{figure}[H]
    \centering
    \includegraphics[width=1\columnwidth]{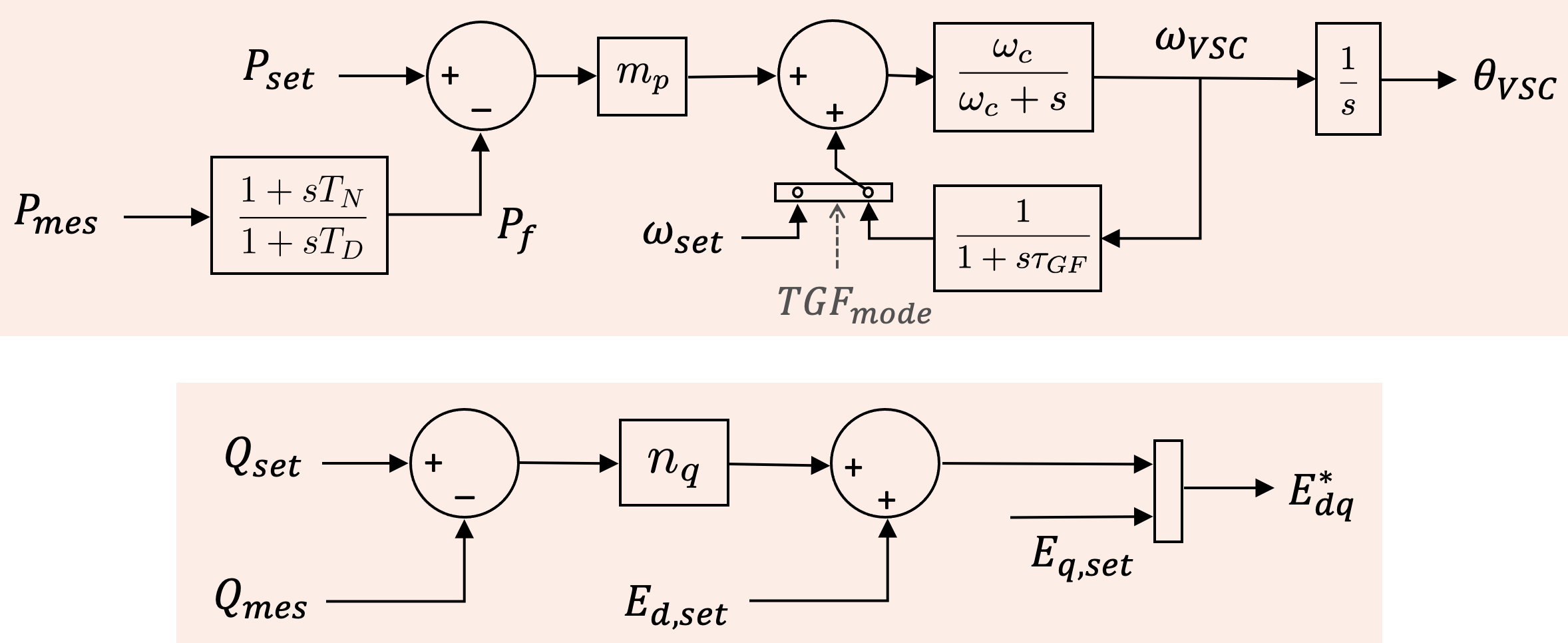}
    \caption{Droop-based active and reactive power controls for grid-forming VSC.}
    \label{fig:model_vsc}
    \vspace{-1em}
\end{figure}

Regarding the specific implementation of the grid-forming control, in this work we consider a filtered droop with a threshold virtual impedance (TVI) for current limitation~\cite{MigrateD32} and a lead-lag in the active power measure~\cite{PrevostACDC2019}. Therefore, the control consists of cascaded current and voltage inner loops whose references are generated by droop-based control laws applied to active and reactive powers, with respective droop coefficients $m_p$ and $n_q$ (see Fig.~\ref{fig:model_vsc}). The injected active and reactive powers in steady state depend on the voltage and frequency according to those settings. All the controls are implemented in per unit.
In this figure, $P_{set}$ ans $Q_{set}$ are the active and reactive power set points. Moreover, the active power error between the set point and the filtered measured power $P_{f}$ is low-pass filtered with a cut-off frequency $\omega_c$. This parameter, together with $m_p$, determine the active power dynamics.
In addition, to decouple the inertial response from the frequency regulation services, we have added an outer loop that brings the VSC active power to its reference by updating the converter frequency reference with a settable dynamics (low-pass filtering of $\omega_{vsc}$ with time constant $\tau_{GF}$). If this $TGF_{mode}$ is not enabled, the reference angular frequency is simply set to $\omega_{set}$. 

\subsection{Equivalent sizing} 
\label{sec:eqsize}
First of all, the nominal apparent power of both devices is set to the same value $S_n$. Second, by comparing equation~\eqref{eq:Hsc} and the control law in Fig.~\ref{fig:model_vsc}, we see that the parameters of the filtered droop control for the VSC can be set to fit the physical characteristic of the SC in terms of~inertia. Indeed, the \textit{equivalent inertia} provided by the VSC, $H_{vsc}$, can be defined as in~\eqref{eq:Heq}. For the sake of comparison a realistic inertia constant is considered for the SC. Then, the product of $\omega_c$ and $m_p$ is fixed accordingly.
\begin{equation}
    H_{vsc} = \frac{1}{2m_p\omega_c}
    \label{eq:Heq}
    \vspace{-2mm}
\end{equation}

Third, the short-circuit current of the SC can be defined as the value at 100 ms, usually named $I_b$, which is associated to the electromagnetic design of the synchronous machine, more specifically its transient impedance $X_d^'$ and the nominal voltage $V_{n,ph}$: 
\begin{equation}
    I_{b,SC} = \frac{V_{n,ph}}{X_d^'}
    \label{eq:Icc}
    \vspace{-2mm}
\end{equation}

However, for the VSC this value is often close to the nominal one and depends strictly on the overcurrent capability of the converter. In this work we do not consider any specific oversizing to match the short-circuit current of the SC. This will be an acknowledged bias in the comparison and we focus on their respective system strength provision. In addition, the devices damping contribution depends on the mechanical and electromagnetic design - including damper winding - for the SC, and on the control tuning - including inner loops~\cite{QoriaPSCC2019} - for the VSC. The mechanical damping is neglected for the SC. 
The factor $K_d$ is given by $\frac{1}{m_p}$ for the VSC. No effort was devoted to match the oscillatory response so different performances are observed.  

\subsection{Equivalent grid model}
\label{sec:grid}
\begin{figure}[H]
    \centering
    \includegraphics[width=0.7\columnwidth]{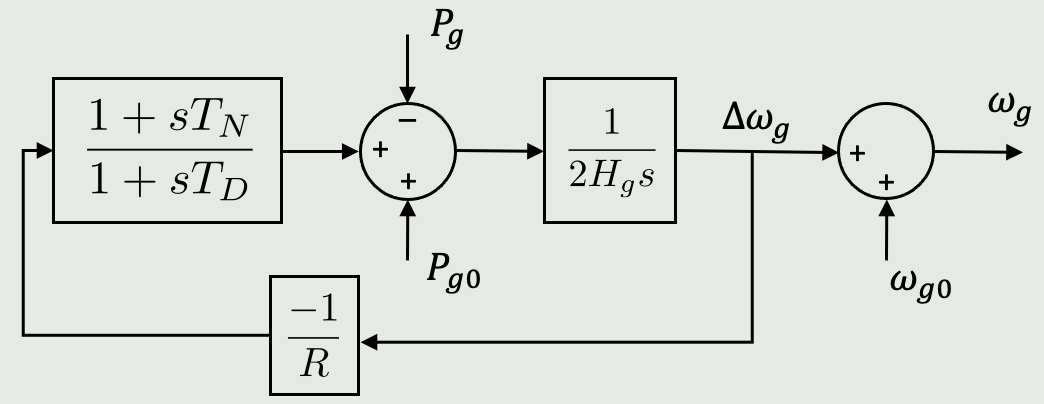}
    \caption{Inertial frequency-response model for equivalent grid.}
    \label{fig:model_grid}
    \vspace{-1em}
\end{figure}

The grid is modeled as an inertial three-phase AC voltage source of variable frequency, $\omega_g$, and controlled amplitude $V_g$. The grid frequency variation, $\Delta \omega_g$, represents the response of the synchronous generators connected to the simulated grid and depends on the aggregated inertia $H_g$. An equivalent droop characteristic represents the primary frequency regulation contribution of the bulk system as shown in Fig.~\ref{fig:model_grid}, based on the active power and grid frequency set points, $P_{g0}$ and $\omega_{g0}$, and the measured active power $P_{g}$. Hence, the grid response consists of:
\begin{itemize}
    \item an integrator with gain $1/(2H_g)$,
    \item a lead-lag $\frac{1+sT_N}{1+sT_D}$ filtering the power deviation, and
    \item a static droop gain $1/R$ on the frequency deviation.
    \vspace{-2em}
\end{itemize} 

The voltage source is placed behind a series impedance $R_g-L_g$ characterizing the grid strength. These values can be defined as a function of the size of the grid-forming devices (i.e. the short-circuit ratio, SCR).

\section{Simulation results}
\label{sec:simu}

Although SC applications are generally of larger capacities, we consider here the following sizing\cite{OSMOSED32}: 
\begin{itemize}
    \item Nominal apparent power : $S_n = 1 \: [MVA]$
    \item Nominal voltage : $U_n = 606 \: [V]$
    \item Nominal frequency : $f_n = 50 \: [Hz]$
\end{itemize}

The SC nominal values are defined accordingly. Then, the inertia constant of the SC is defined as $H_{SC}=$~5~s. The parameters of the grid-forming control are set to $m_p=$ 0.04 and $w_c=$ 2.5 rad/s in order to obtain the same equivalent inertia. The considered SC electromagnetic design entails a short-circuit current of almost 6 pu\footnote{$X_d=$2.24, $X_d^{'}=$0.17, $X_d^{''}=$0.12, $X_q=$1.02, $X_q^{''}=$0.13, $X_I=0.08$, $R_s=$ 0.017 pu,  $T_{d0}^'=$ 4.4849 s, $T_{d0}^{''}=$ 0.0681 s, $T_q^{''}$=0.1 s}, while the VSC overcurrent capability is set to 10\%. The VSC output filter inductance is 0.15 pu. Finally, the transformer reactance is also set to 0.15 pu. 

On the grid side, we consider a SCR of 3 ($L_g=0.33, R_g=0.033$ in $S_n$ base), 
$H_g=$ 5 s, $R=$ 4\%, $T_N=$ 1 and $T_D=$ 6. Here $H_g$ is defined with respect to the system size $S_{n,grid}=SCR * S_n$. Finally $P_{g0}$ is set to the initial load which is 0.5 pu in $S_{n,grid}$ base.


\subsection{Simulation conditions}

The VSC reference power $P_{set}$ is set to zero to have initial conditions close to the ones of the SC. The time constant of the active power restoration loop is set to $T_{GF}=$ 100 ms in order to match the SC active power response after some seconds. The reactive power droop is initially set to zero ($n_q = 0$) to regulate the voltage at the same reference value as the SC. 

The transient behaviour of the VSC and SC are compared by overlapping the instantaneous active and reactive power, the frequency, the voltage and current. For the voltage and current, we display the norm of the values in pu, multiplied by the nominal RMS values. We consider three tests:
\begin{itemize}
    \item A load variation $\Delta P_{load} = -0.4$ p.u (base $S_{n,grid}$).
    \item A grid voltage variation $\Delta V_{grid} = -5\%$
    \item A 150 ms three-phase fault at the point of common coupling (PCC, the primary side of the transformer).
\end{itemize}

\subsection{Inertial response for load variation}

\begin{figure}[H]
    \centering
    \begin{subfigure}[b]{0.49\columnwidth}
        \centering \includegraphics[width=\columnwidth,trim=0 0 15cm 11cm, clip]{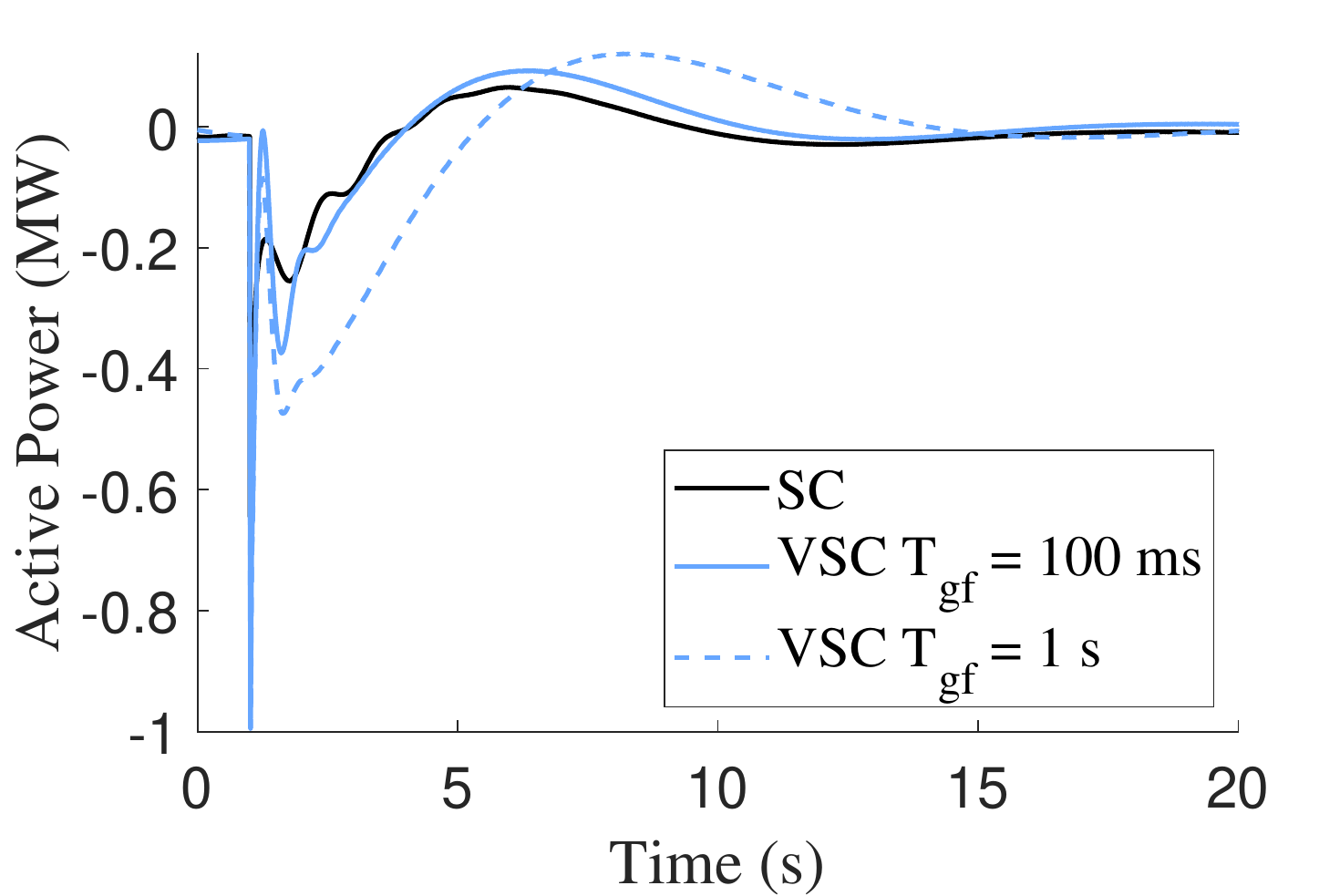}
        \caption{Active power at PCC}
    \end{subfigure}
    \begin{subfigure}[b]{0.49\columnwidth}
        \centering \includegraphics[width=\columnwidth,trim=0 0 15cm 11cm, clip]{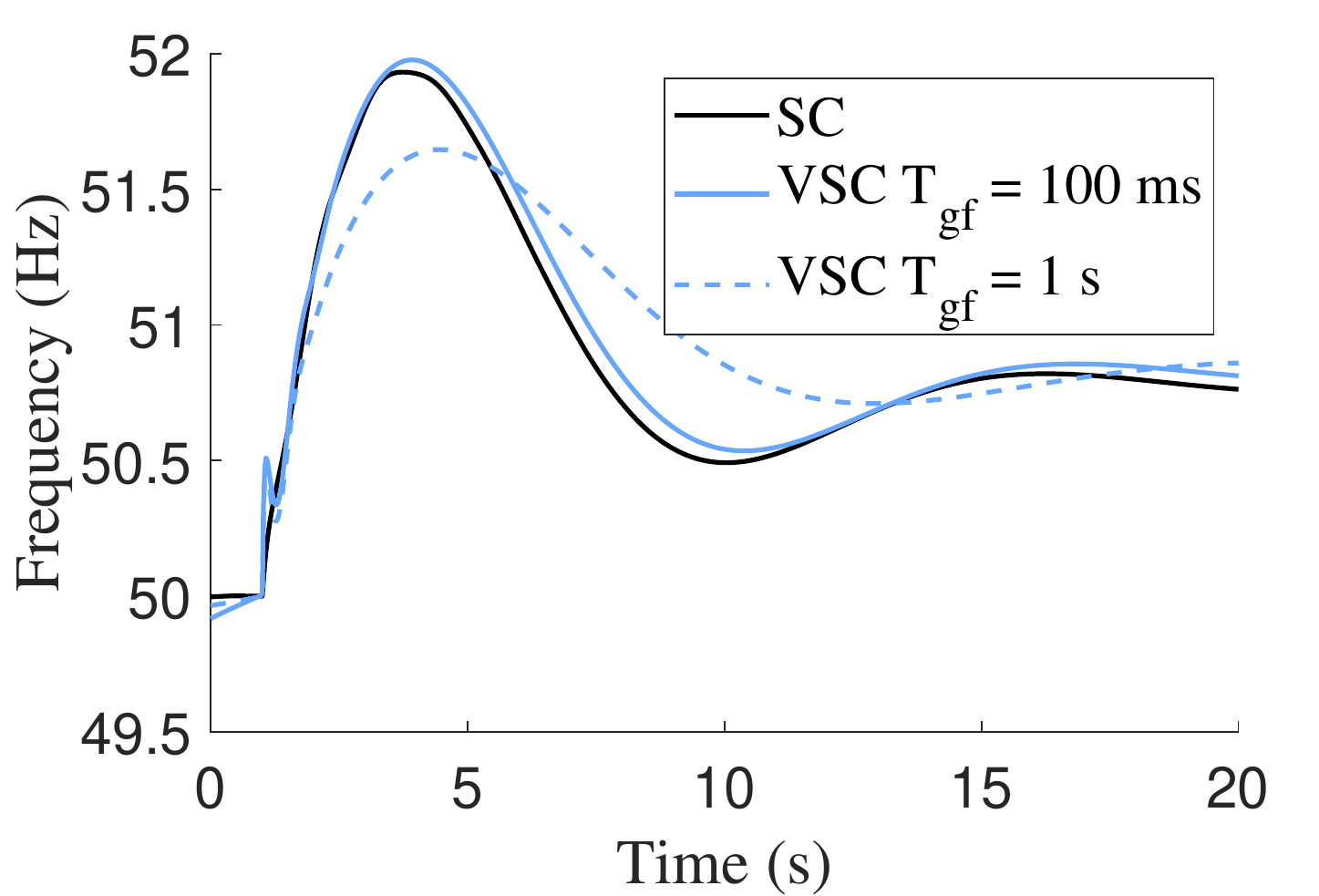}
        \caption{Frequency of the device}
    \end{subfigure}
    
    \begin{subfigure}[b]{0.49\columnwidth}
        \centering \includegraphics[width=\columnwidth,trim=0 0 15cm 11cm, clip]{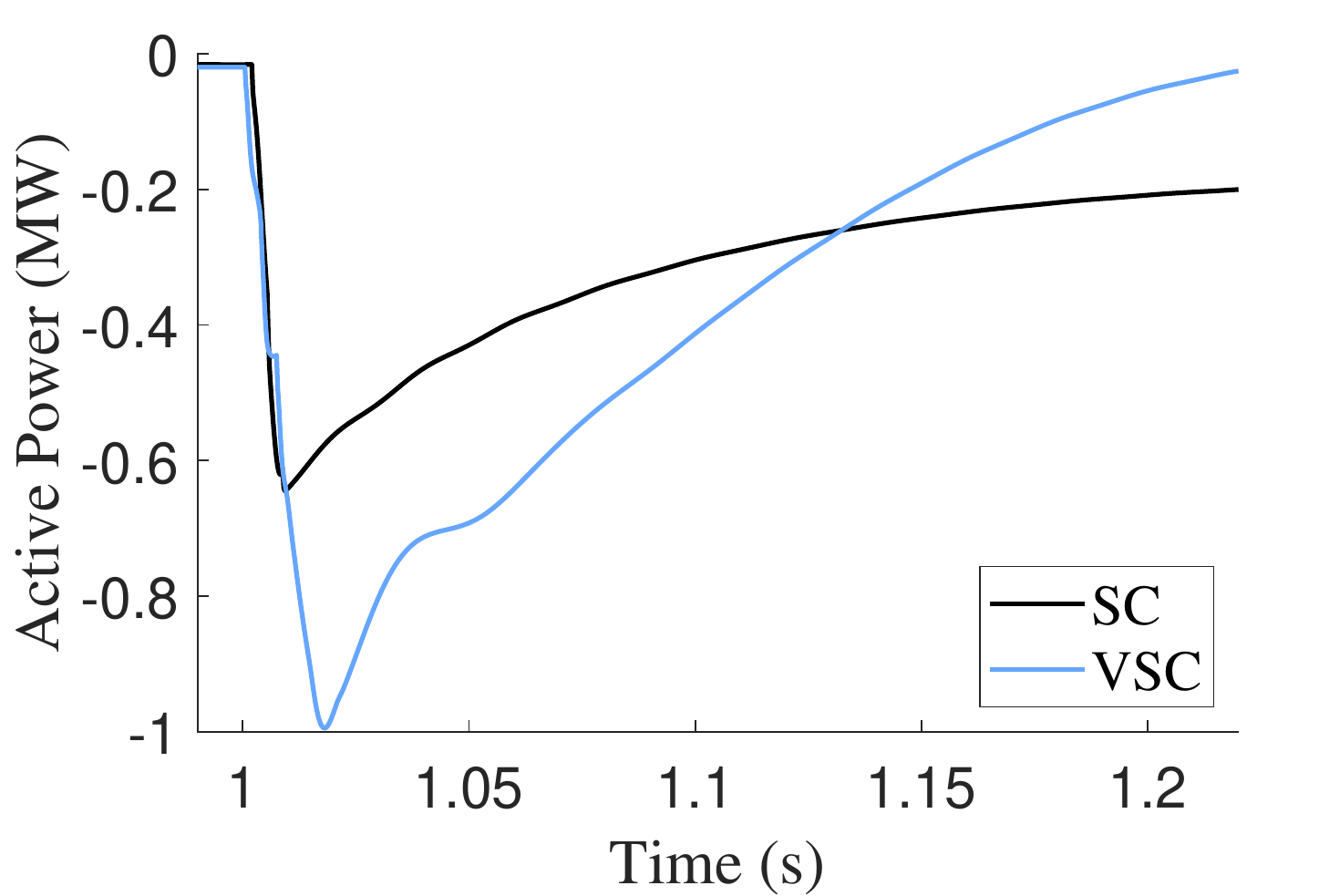}
        \caption{Active power at PCC (zoom)}
    \end{subfigure}
    \begin{subfigure}[b]{0.49\columnwidth}
        \centering \includegraphics[width=\columnwidth,trim=0 0 15cm 11cm, clip]{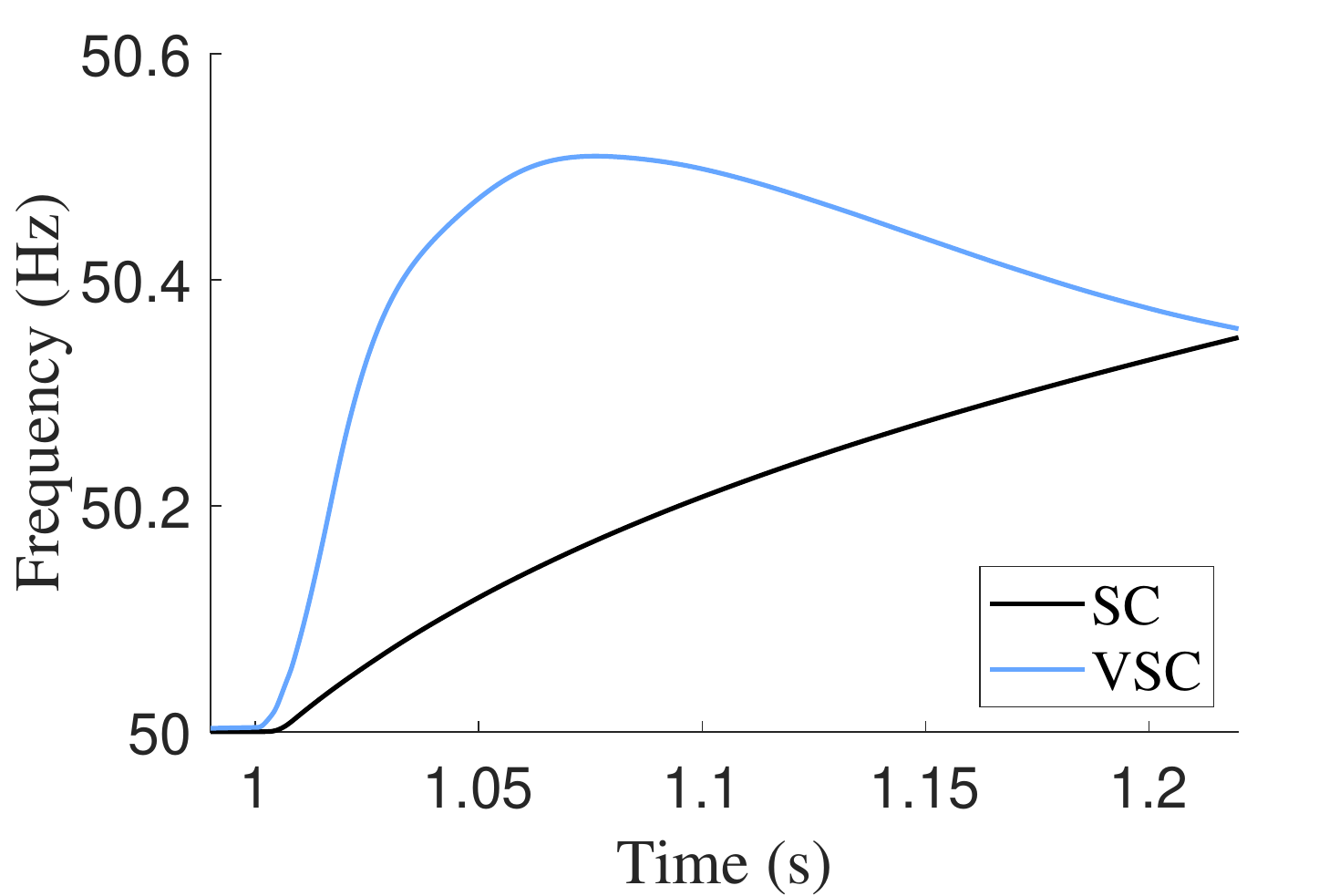}
        \caption{Frequency of the device (zoom)}
    \end{subfigure}
    
\caption{VSC vs SC responses to a load disconnection.}
\label{fig:load}
\vspace{-1em}
\end{figure}

Figure~\ref{fig:load} confirms that a grid-forming VSC can be tuned to provide the same instantaneous active power contribution as the one given by a SC for load variations. Moreover, by reducing the cut-off frequency of the low pass filter ($\omega_{GF}=1/T_{GF}$) used for reference tracking in the grid-forming VSC, the power boost can be increased and maintained for a longer period to provide additional system support, limiting further the frequency excursion. 
However a closer look to the first 200 ms reveals that the VSC does experience a slightly higher initial rate of change of the frequency ($ROCOF$) during 50 ms, which is then quickly compensated to catch up the SC response. As $ROCOF$ measurements are generally based on 200-500 ms averages, this phenomenon should have limited impact on neighbouring devices relying on this variable.
\vspace{-0.5cm} 

\subsection{Device strength for a grid voltage variation}
\begin{figure}[H]
    \centering
    \begin{subfigure}[b]{0.49\columnwidth}
        \centering \includegraphics[width=\columnwidth,trim=0 0 15cm 11cm, clip]{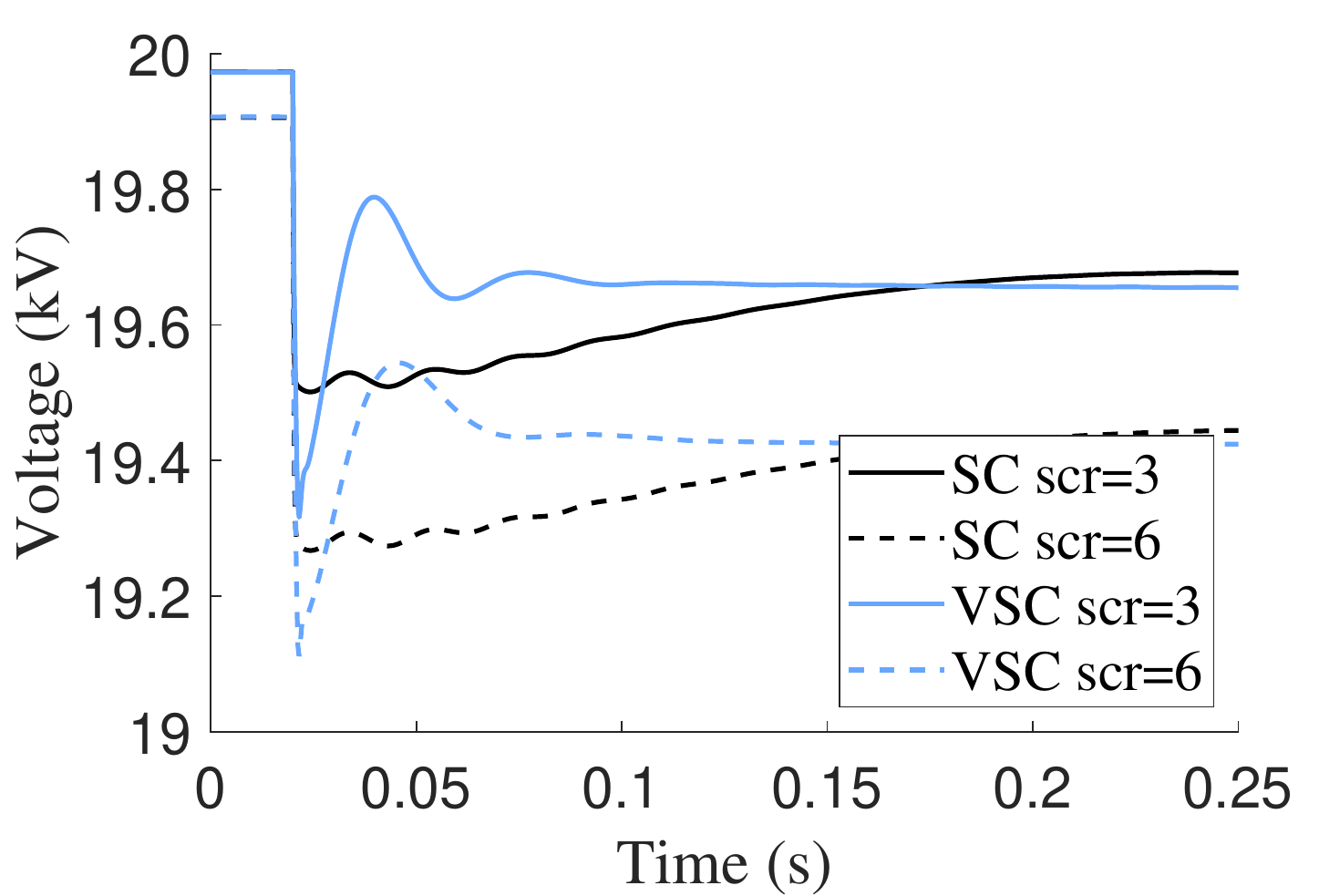}
        \caption{Voltage at PCC}
    \end{subfigure}
    \begin{subfigure}[b]{0.49\columnwidth}
        \centering \includegraphics[width=\columnwidth,trim=0 0 15cm 11cm, clip]{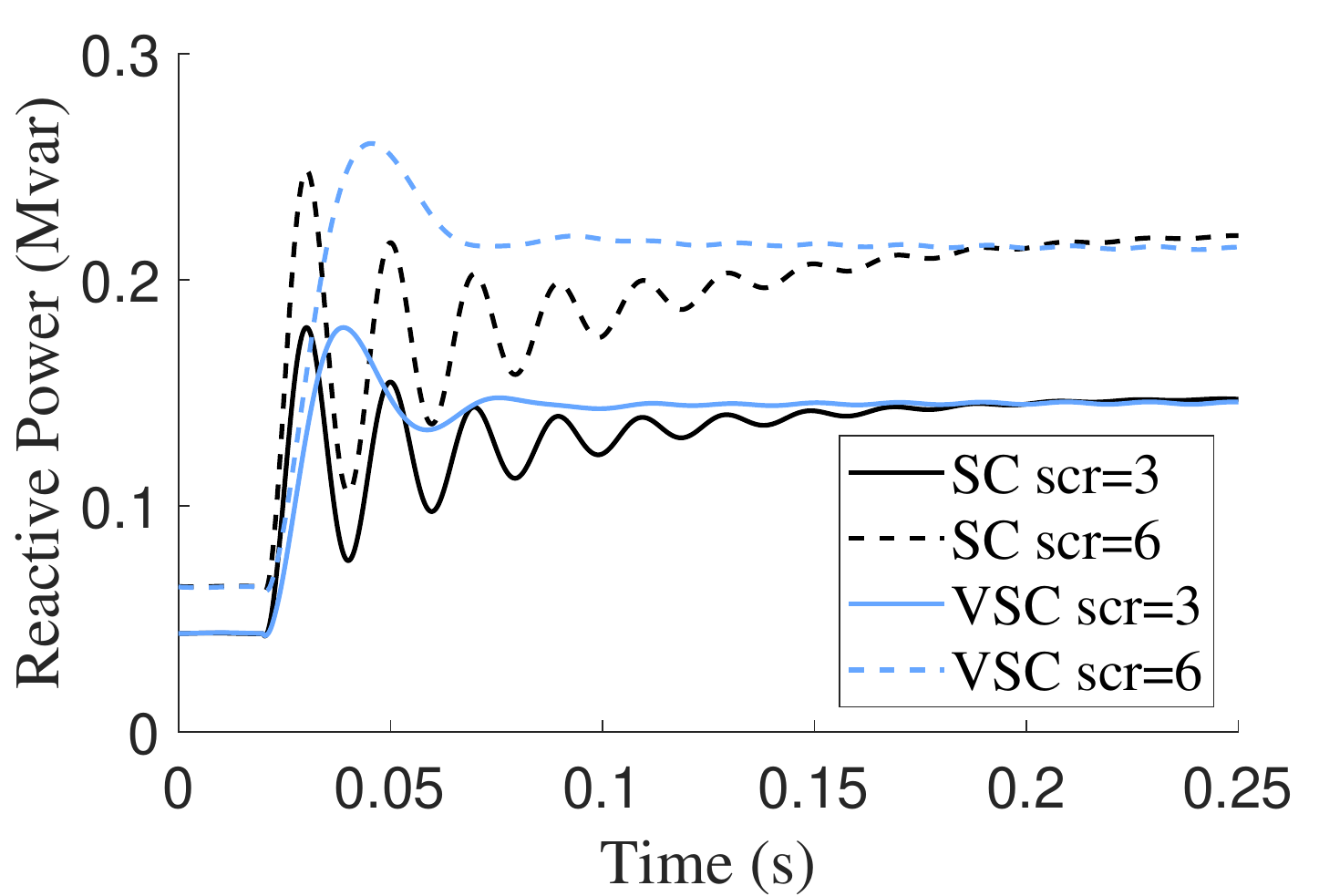}
        \caption{Reactive power at PCC}
    \end{subfigure}
\caption{VSC vs SC responses to a 5\% grid voltage decrease.}
\label{fig:volt}
\vspace{-3mm}
\end{figure}

A 5\% voltage decrease on the controlled source that represents the grid ($V_{g}$) entails a voltage variation at PCC, which amplitude depends on the SCR and the load but also on the strength contribution of the device under test. For illustrative purposes, we also include simulation results with a stronger grid (SCR=6). Figure \ref{fig:volt} shows that both the SC and the VSC instantaneously inject reactive power following a voltage dip and provide the same contribution after around 100-150 ms depending on SCR. Nevertheless during the first cycle the SC exhibits a faster response limiting the voltage drop at PCC to 2\% in the base case (SCR=3) and to 2.5\% when the SCR was increased to 6, while the grid-forming VSC allows the voltage to drop up to 3.5\% and 4\% respectively. Between these two time scales (10-100 ms), the VSC shows as expected a faster and better damped response. However, the SC voltage regulator could be improved, including power system stabilizer (PSS), to offer performances closer to the VSC ones. 
Finally, we acknowledge that opposite results could appear in operating conditions near to the VSC stability limits.
\vspace{-0.3cm} 

\subsection{Short-circuit current supplied in a three-phase fault}

\begin{figure}[H]
    \centering
    \begin{subfigure}[b]{0.49\columnwidth}
        \centering \includegraphics[width=\columnwidth,trim=0 0 15cm 11cm, clip]{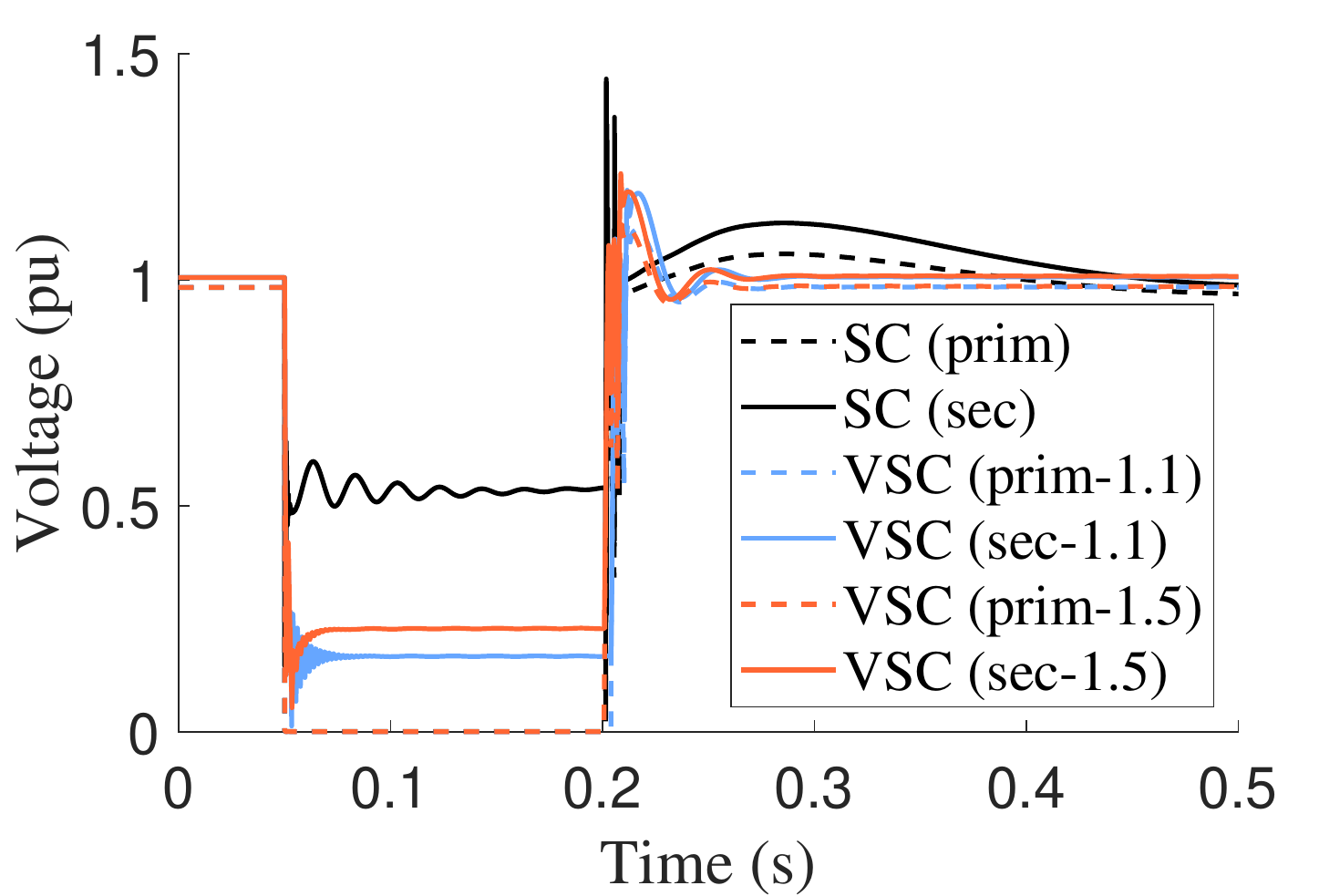}
        \caption{Voltage at PCC}
    \end{subfigure}
    \begin{subfigure}[b]{0.49\columnwidth}
        \centering \includegraphics[width=\columnwidth,trim=0 0 15cm 11cm, clip]{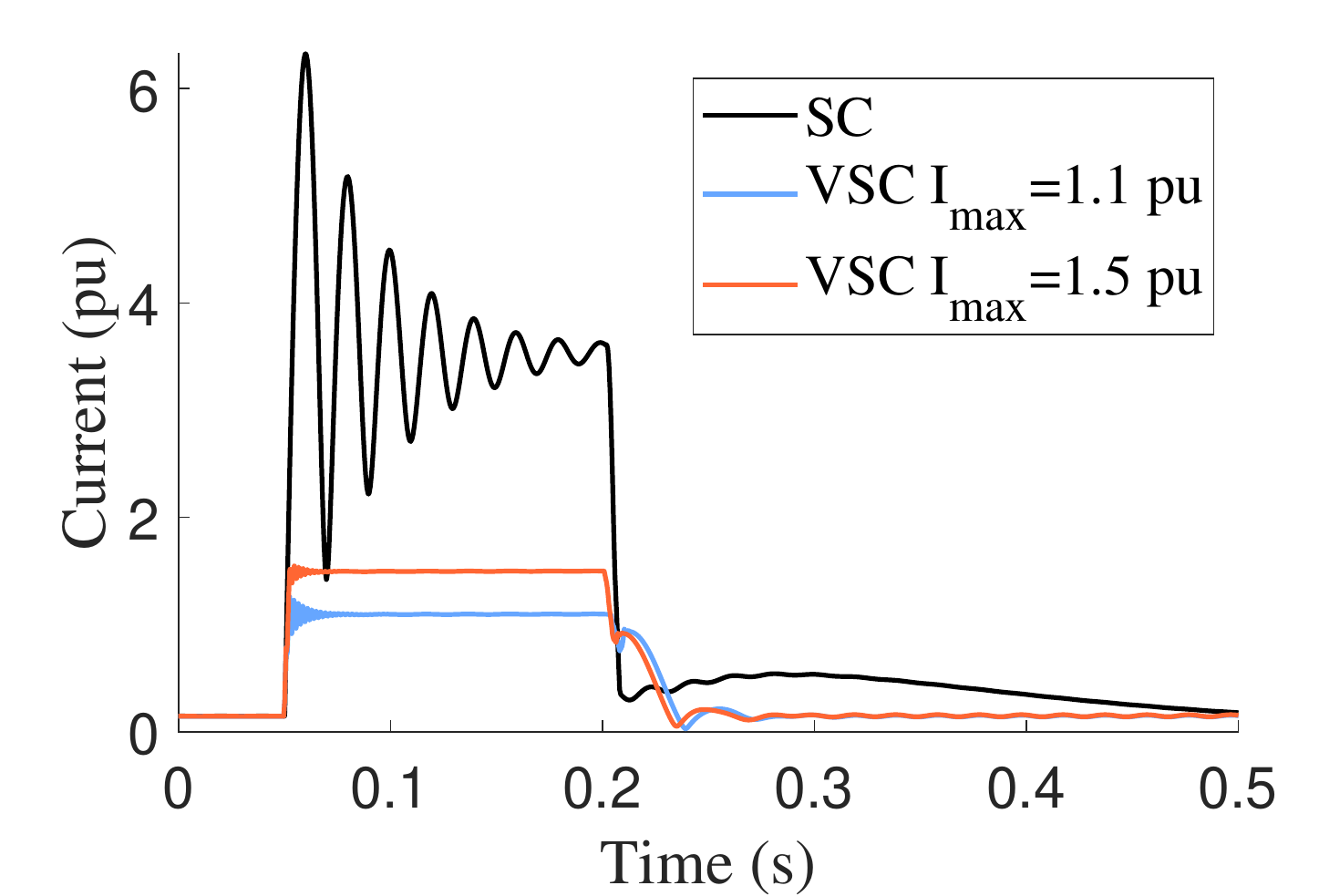}
        \caption{Current at PCC}
    \end{subfigure}
    
    \centering
    \begin{subfigure}[b]{0.49\columnwidth}
        \centering \includegraphics[width=\columnwidth,trim=0 0 15cm 11cm, clip]{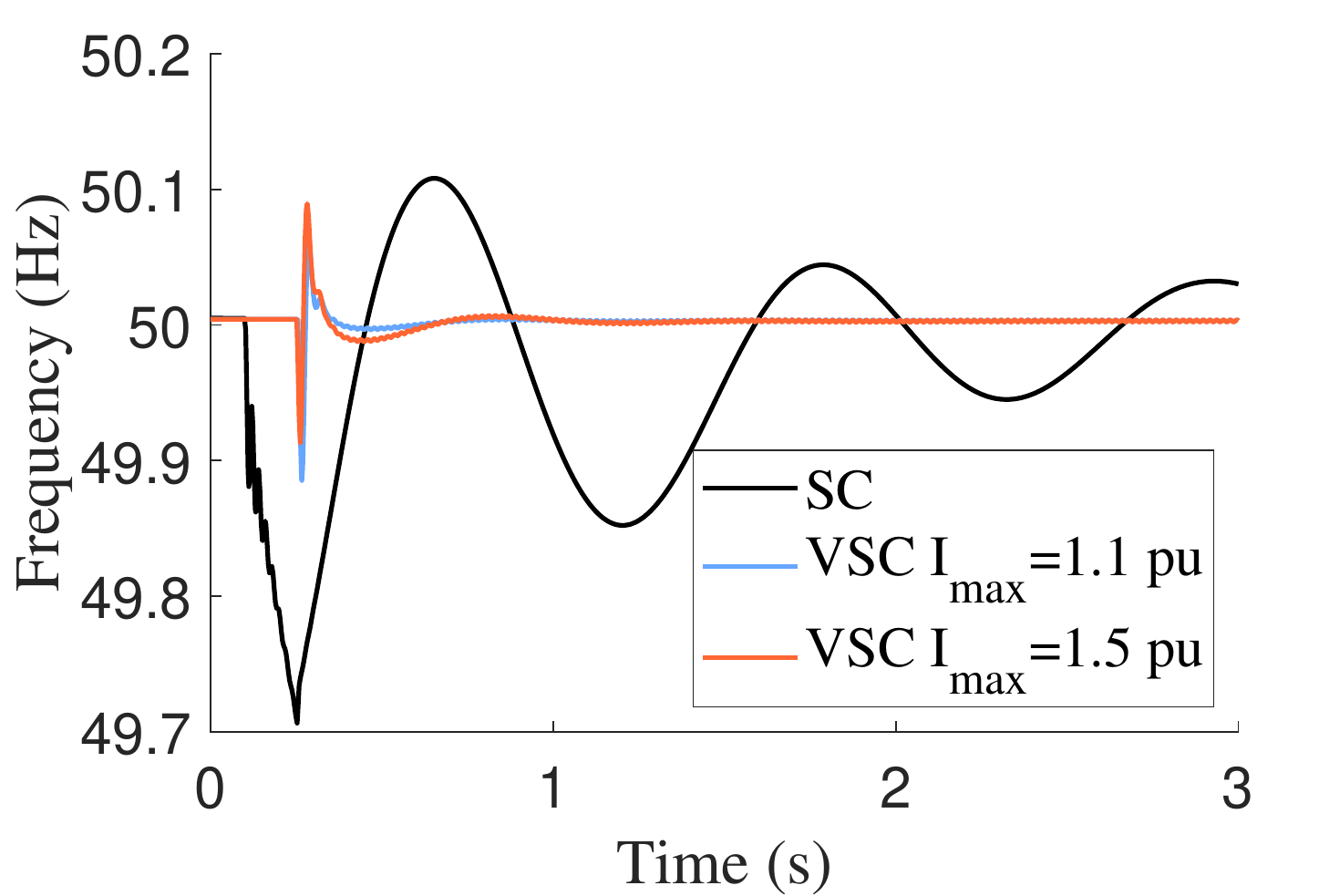}
        \caption{Frequency of the device}
    \end{subfigure}
    \begin{subfigure}[b]{0.49\columnwidth}
        \centering \includegraphics[width=\columnwidth,trim=0 0 15cm 11cm, clip]{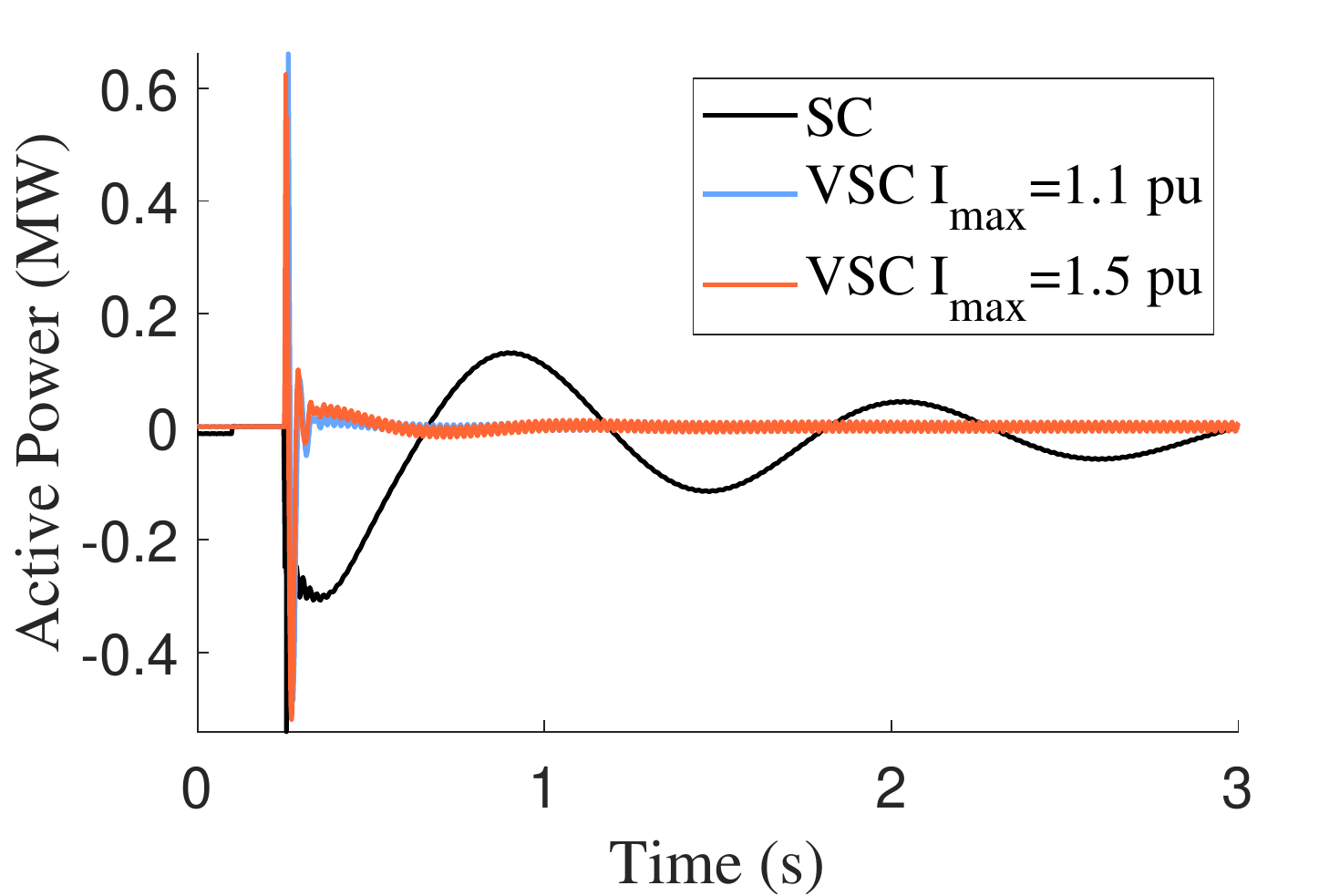}
        \caption{Active power}
    \end{subfigure}
\caption{VSC vs SC responses to a three-phase fault.}
\label{fig:3ph}
\end{figure}

Figure~\ref{fig:3ph} shows that both devices comply with the selected fault ride through (FRT) profile, instantaneously inject fault current and successfully synchronize after the fault clearance. 
As expected, during the fault, the SC provides a short-circuit current that rises above 6 p.u and decays to 3.5 pu after 150 ms, while the grid-forming VSC contribution is directly limited by its overcurrent capability. We show the voltage at both sides of the transformer in pu and we present two cases for the VSC: the base case (blue curve) has a current threshold for TVI activation at 1 pu and a maximal current of 1.1 pu; and second case has a TVI activation threshold of 1.25~pu and a maximal current of 1.5~pu (orange curve).

The higher the current, the higher is the residual voltage at the secondary side of the transformer. Therefore, the SC offers a better voltage support (the amplitude at the low voltage side is maintained to 0.5 pu). However, the SC exhibits some power oscillation with a frequency excursion during the fault, while the VSC frequency remains close to the nominal value for the test case ($P_{ref}=0$). A transient is only observed during the resynchronisation with the main grid at fault clearance. 
The SC oscillations on the frequency and active power take about 2 seconds to get damped. 
Therefore, although the SC provides higher short-circuit current, the VSC 
resynchronizes and recovers reference tracking faster. 

\subsection{Discussion on services and sizing}
\label{sec:discussion}

The selected tests illustrate that after 200 ms the grid support provided by the grid-forming VSC and the SC are equivalent. At this time scale, fast active and reactive power controls in grid-following converters could allow similar performances. Here we focus on the behaviour between the first cycle and 100 ms \footnotemark. 

\subsubsection{System strength}

\footnotetext{Assessing the differences below 1 cycle would require more detailed models than the ones considered in this work. In addition, they would be hard to exploit as a discriminating criteria in operation.}

The grid-forming control gives the VSC the capability to provide the same contribution to the system strength as the SC, initiating reactive power injection immediately after grid disturbances. This can be achieved without adding any additional equipment or oversizing by prioritizing current limitation. TSO can manage to maintain a reactive power reserve on grid-forming units for this purpose through a secondary voltage control. 
In the proposed example, the VSC outperforms the SC in terms of voltage stiffness between 10-100 ms. 
The SC seems to outperform the VSC within the first cycle, which can be partially explained by the lower sub-transient impedance $X_d^{''}$ (for SC) compared to the considered VSC output filter impedance. However, this preliminary result is at the validity limits of the selected models and therefore, requires further investigation. 

\subsubsection{Inertial response}
Analogously, the grid-forming control can be tuned to provide an electrical inertia equivalent to the physical inertia of the SC, leading to a very similar instantaneous active power injection. It is worth highlighting that this parameter will be fixed for the SC, while it can be reconfigured for the VSC, in order to adapt to the evolution of the system needs and potentially adjust the installation instantaneous capabilities. Again, if the response brings the converter to its current limit, the TVI can curtail the installation contribution to avoid oversizing. However, this function raises the question of the energy storage capabilities. This is not an issue for a BESS that can be overloaded for few seconds. Other resources, such as wind turbines, provide naturally some energy storage suitable for this application, but that will depend on the operating point. At low wind speed it might not be possible to extract any energy from the rotating masses. Ensuring a firm provision of this service by different resources, including PV plants for instance, might lead in some cases to install additional equipment.

\subsubsection{Short-circuit current}
The overcurrent capability of commercial VSC varies from one manufacturer to the other, but it is definitely way below the SC. VSC sizing is therefore very sensitive to this requirement. The specific system needs must be precisely quantified and the most suitable solution be determined by a technical-economic analysis.

\newpage
\section{Conclusion}
\label{sec:conclusion}

The replacement of SG by RES connected to the grid through grid-following converters entails a reduction of the system inertia and strength, making the grid voltage frequency and amplitude more volatile. This phenomenon is leading to stability issues worldwide in system concentrating high share of RES in specific areas. New system needs have been defined to ensure minimal voltage stiffness (in amplitude and frequency), which can be fulfilled by dedicated devices installed by TSO or by grid-connected resources (generators, FACTS, HVDC links...) through CNC requirements or service provision.

The installation of SCs has proved to be an effective solution. In addition, they provide high short-circuit current that can also be required in some cases to ensure proper operation of protection schemes. Other than the investment cost, this solution entails operational costs linked to the losses (friction, overheating, circuit losses) that range between 1\% and 1.5\% of the rated power~\cite{ref20}.

This work has shown that these functions can be performed, to some extent, by a standard VSC with a grid-forming control. The VSC with adequate controls and the SC provide equivalent services in terms of contribution to system strength and inertia. The first one could be established as a minimal requirement, within the converter current capability. The second could be requested as constructive capability with an effective contribution that depends on the intrinsic storage capability of the installation (within a range of $H$ for instance), exploiting the fact that the VSC response is defined by control parameters, instead of physical properties. This should rule out hardware upgrades (oversized equipment or installation dedicated to storage) at this first level of specification. However, high current requirement could prevent the deployment of VSC in favor of SC, as the necessary oversizing would greatly increase the costs.



\section*{Acknowledgment}
This work is part of the OSMOSE project. The project has received funding from the European Union’s Horizon 2020 research and innovation program (grant agreement No 773406). This article reflects only the authors views and the European Commission is not responsible for any use that may be made of the information it contains.

\end{document}